\documentstyle[aps,preprint]{revtex}

\tighten
\begin{document}
\draft

\title{Higher-order binding corrections to the Lamb shift of $2P$ states}

\author{U. D. Jentschura$\dag$ and K. Pachucki$\ddag$}

\address{
$\dag$ Max--Planck--Institut f\"{u}r Quantenoptik,
Hans-Kopfermann-Stra\ss e 1,
85748 Garching, Germany\cite{emails}}

\address{
$\ddag$ Institute of Theoretical Physics,
Warsaw University,
Ho\.{z}a 69,
00-681 Warsaw, Poland \cite{emails}}      

\maketitle

\begin{abstract}
We present an improved calculation 
of higher order corrections to the one-loop self energy of 
$2P$ states in hydrogen-like systems with small nuclear charge $Z$.
The method is based on a division of the integration with respect to
the photon energy into a high and a low energy part. 
The high energy part is calculated by an expansion of
the electron propagator in powers of the Coulomb field. The low energy 
part 
is simplified by the application of a Foldy-Wouthuysen transformation. 
This transformation leads to a clear separation 
of the leading contribution from the relativistic 
corrections and removes higher order terms.
The method is applied to the $2P_{1/2}$ and $2P_{3/2}$ 
states in atomic hydrogen. The results lead to new theoretical 
values for the Lamb shifts and the fine structure splitting.
\end{abstract}

\narrowtext

\section{Introduction}
The evaluation of the one-loop self-energy of a bound electron
is a long standing problem in Quantum Electrodynamics.
There are mainly two approaches. The first, developed by P. Mohr 
\cite{mohr2},
relies on a multidimensional numerical
integral involving a partial wave expansion of
the electron propagator in the Coulomb field.
This approach is particularly useful for  
heavy hydrogen-like ions. The second approach
is based on an expansion of the electron 
self-energy in powers of $Z\,\alpha$,
\begin{equation}
\delta E_{\rm SE}  =   \frac{\alpha}{\pi} \, (Z \alpha)^4 \, m \, F\,,
 \label{definitionofF}
\end{equation}
where
\begin{eqnarray}
F &=& A_{40} + A_{41} \, \ln\left[(Z \alpha)^{-2}\right] + 
  (Z \alpha) \, A_{50} +\nonumber \\
& &  (Z \alpha)^2 \left(A_{60} + 
A_{61} \,\ln\left[(Z \alpha)^{-2}\right] + 
   A_{62} \,\ln^2\left[(Z \alpha)^{-2}\right] + o(Z \alpha) \right).
\label{2}
\end{eqnarray} 
The leading contribution as given by $A_{41}$ has been originally
calculated by Bethe in \cite{bethe}. Many others have contributed
to the evaluation of higher orders corrections, for details see 
an excellent review by Sapirstein and Yennie in \cite{sapirstein}.
A very general analytical method has been introduced
by Erickson and Yennie in \cite{erick}. Erickson and Yennie were able to
calculate all the coefficients in (\ref{2}) except for $A_{60}$.
The calculation of corrections of $(Z\,\alpha)^2$ relative order
is a highly nontrivial task because the binding Coulomb field enters
in a nonperturbative way, and there is no closed form expression
for the Dirac-Coulomb propagator. Additionally, one-loop
electron self-energy contributes to all orders in $Z\,\alpha$,
and the separation of the $(Z\,\alpha)^2$ relative contribution
involves hundreds of terms.
A very efficient scheme of the calculation has been introduced in 
\cite{krpdissertation}.
It was calculated there the $A_{60}$ coefficient for the $1S$ and $2S$ 
states
in hydrogen atom. The method was based on the division
of the whole expression into two parts, $E_L$ and $E_H$,
by introducing an artificial parameter $\epsilon$
which  is a cutoff in the photon frequency.
In the high energy part $E_H$ one expands the electron propagator
in powers of the Coulomb field and uses Feynman gauge. 
In the low energy part
one uses Coulomb gauge and applies a multipole expansion.
The most important ingredient of this method
is the expansion in the parameter $\epsilon$
after the expansion in $Z\,\alpha$ is performed
(for details see the next section).

The calculation presented in this paper is a further 
development of this original method. In the low energy part we
use a Foldy-Wouthuysen transformation. The transformation clearly 
identifies the leading order contribution and separates out
all higher order terms. An additional advantage is that the
nonrelativistic Schr\"{o}dinger-Coulomb propagator can be used here.
A closed-form expression of this propagator is known
in coordinate and in momentum space (for details see 
\cite{swainsondrake}). 
This method is applied to  
the $2P_{1/2}$ and $2P_{3/2}$ states. All coefficients including 
$A_{60}$ are obtained. We recover all the previously known
results, and the new results for $A_{60}$ are in agreement with those
obtained from the extrapolation of P. Mohr's data. Our results are 
relevant for single electron, small $Z$ systems 
(for example atomic hydrogen and ${\rm He}^{+}$), which are currently
investigated with very high precision. 
New theoretical values for the Lamb shift of the $2P_{1/2}$ and 
$2P_{3/2}$ states and the fine structure summarize our calculations.

\section{The $\epsilon$\lowercase{fw}-method} 

The self-interaction
of the electron leads to a shift of the hydrogen energy levels.
This shift at the one-loop level is given by 
\begin{equation}
\label{deltaESE}
\delta E_{\rm SE}= i e^2 \int \frac{d^4 k}{(2 \pi)^4} 
  D_{\mu \nu}(k) \langle \bar{\psi} | \gamma^{\mu} 
  \frac{1}{\not\!p - \not\!k - m - \gamma^0 V} \gamma_{\nu} 
  | \psi \rangle - \langle \bar{\psi} | \delta m | \psi \rangle,
\end{equation}
where $\delta m$ refers to the mass counter term, and it is 
understood that the photon propagator $D_{\mu \nu}$ has to be
regularized to prevent ultraviolet divergences. $\bar{\psi}$ 
is the Dirac adjoint $\bar{\psi} = \psi^{+} \, \gamma^0$.

\begin{figure}[htb]
\begin{center}
\setlength{\unitlength}{0.015in}
\begin{picture}(280,100)(-140,-50)
\thicklines
\put (-120,0){\line(1,0){272}}
\put (-.4,-45){\line(0,1){92}}
\thicklines
\put (151,0){\vector(1,0){4}}
\put (-.4,46) {\vector(0,1){4}}
\thinlines
\put (2,-3){\line(1,0){135}}
\put (-2,3){\line(-1,0){113}}
\put (80,-6){\line(1,0){57}}
\put (-24,6){\line(-1,0){91}}
\put (17,0){\oval(34,34)[tl]}
\put (17,17){\line(1,0){124}}
\put (-17,0){\oval(34,34)[br]}
\put (-17,-17){\line(-1,0){102}}
\thicklines
\put (25,0){\oval(50,50)[l]}
\put (25,25){\line(1,0){117}}
\put (25,-25){\line(1,0){117}}
\put (75,25){\vector(1,0){4}}
\put (80,-25){\vector(-1,0){4}}
\put (75,17){\vector(1,0){4}}
\put (-75,-17){\vector(1,0){4}}
\thinlines
\put (50,-3){\line(0,1){6}}
\put (50,-28){\line(0,1){6}}
\put (50,22){\line(0,1){6}}
\put (80,-3){\line(0,1){6}}
\put (-6,4.7){${\scriptstyle{\times}}$}
\put (-12,4.7){${\scriptstyle{\times}}$}
\put (-17,4.7){${\scriptstyle{\times}}$}
\put (-21,4.7){${\scriptstyle{\times}}$}
\put (-24,4.7){${\scriptstyle{\times}}$}
\put (135,5){$\Re(\omega)$}
\put (-27,40){$\Im(\omega)$}
\put (15,5){$C_F$}
\put (-50,-12){$C_F$}
\put (15,30){$C_L$}
\put (15,-20){$C_L$}
\put (90,30){$C_H$}
\put (90,-20){$C_H$}
\put (51,3){$\epsilon$}
\put (81,3){$2\,m$}
\end{picture}
\end{center}
\caption{\label{intcontour} The $\omega$-integration contour used
in the calculation. Bending the Feynman contour $C_F$ in the specified way
leads to the high and low energy parts $C_H$ and $C_L$. Lines
directly below and above the real axis denote
branch cuts from the photon and electron propagator. Crosses denote 
poles originating from the discrete spectrum of the electron 
propagator.}
\end{figure}
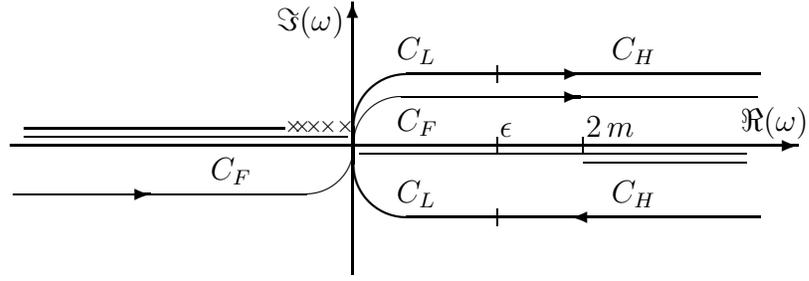

For the $\omega$-integration ($k_0 \equiv \omega$), 
the lower part of the Feynman integration contour $C_F$ is bent 
into the ``right'' half plane with $\Re(\omega) > 0$ and divided into
two parts, the low energy contour $C_L$ and the high energy contour
$C_H$, see Fig. \ref{intcontour}.
The $\epsilon$ parameter corresponds to the cut-off $K$
which was introduced by H. Bethe in his original evaluation of 
the low energy part of the 
electromagnetic shift of energy levels \cite{bethe}
(specifically, $K = \epsilon \, m$).
The two contours are separated along the line $\Re(\omega) = \epsilon \, 
m$,
where $\epsilon$ is some arbitrary dimensionless parameter, 
which we assume to be smaller than unity. 
This method of $\omega$-integration has been described in detail
in \cite{krpdissertation}. 
The two integrations lead to the
high and low energy parts $E_L$ and $E_H$, which are functions of
the fine structure constant $\alpha$ and of the 
free parameter $\epsilon$. Their sum, however,
\begin{equation}
\delta E_{\rm SE}(\alpha) = E_L(\alpha, \epsilon) + E_H(\alpha,\epsilon),
\end{equation}
does not depend on $\epsilon$. The most important step is the expansion in 
$\epsilon$ after the expansion in $\alpha$. It eliminates,
without actual calculations, many terms that
vanish in the limit $\epsilon \rightarrow 0$.
To be more specific, in expanding $E_L$ and $E_H$ in $\epsilon$ we keep 
only 
finite terms (the $\epsilon^0$-coefficients) and the terms which diverge 
as $\epsilon \to 0$. The divergent terms cancel out in the sum,
the finite terms contribute to the Lamb shift.
This cancelation of the divergent terms is an important 
cross-check of the calculation. One may use different gauges of the photon
propagator for the two parts, because the gauge-dependent term 
vanishes in the limit $\epsilon \to 0$.
For convenience, we  use the Feynman gauge for the
high and the Coulomb gauge for the low energy part. 

In this work, 
the treatment of the low energy part is largely simplified by the
introduction of a Foldy-Wouthuysen (fw) transformation.  
It enables one to clearly separate out the leading 
(nonrelativistic dipole) term, which gives the 
$\alpha (Z \alpha)^4$-contribution, from the relativistic corrections,
which give terms in $\alpha (Z \alpha)^6$. An additional advantage 
is the fact that all contributions to the low energy
part can be evaluated using the nonrelativistic 
Schr\"{o}dinger-Coulomb-Green's function, 
whose closed-form solution is well known \cite{swainsondrake}.
Terms which contribute to the Lamb shift up to
$\alpha (Z \alpha)^6$ can be readily 
identified, and each of these can be calculated independently.
In the low energy part we may expand in the photon momentum $k$.
The terms which contribute to the Lamb shift
in the order of $\alpha (Z \alpha)^6$
correspond to the ``non-relativistic dipole'' term 
(involving the non-relativistic 
propagator and  wave function), the ``non-relativistic quadrupole'' term
and the ``relativistic dipole'' term (which involves the relativistic 
corrections to the wave function and the Dirac-Coulomb propagator).
The terms of higher order in $k$ vanishes in the limit 
$\epsilon\rightarrow 0$.

Calculations of the high-energy part are performed almost entirely 
with the computer algebra system {\em Mathematica} \cite{mathematica}.
Because of the presence of an infrared cut-off, one can expand 
the Dirac-Coulomb propagator in powers of the Coulomb potential. 
A subsequent expansion of the propagator 
in electron momenta is also performed.
This leads finally to the calculation of matrix elements of operators
containing $V$ and $\bf{p}$ on the P-states.
Because $P$-wave functions vanish at the origin, all of the
relevant matrix elements are finite up to the order of 
$(Z \alpha)^6$.
 
\section{The high-energy part}

In this part we use the Feynman gauge  
($D_{\mu \nu}(k) = - g_{\mu \nu} / k^2$).
and the Pauli-Villars regularization for the photon propagator 
\begin{equation}
\label{paulivillars}
\frac{1}{k^2} \to \frac{1}{k^2} - \frac{1}{k^2 - M^2},
\end{equation}
so that the following expression remains to be evaluated:
\begin{equation}
E_H = - i e^2 \int_{C_H} \frac{d^4 k}{(2 \pi)^4} 
  \left[ \frac{1}{k^2} - \frac{1}{k^2 - M^2} \right]
   \langle \bar{\psi} | \gamma^{\mu} 
  \frac{1}{\not{p} - \not{k} - m - \gamma^0 V} \gamma_{\mu} 
  | \psi \rangle - \langle \bar{\psi} | \delta m | \psi \rangle
\end{equation}
We start by calculating the matrix element
\begin{equation}
{\tilde P} = \langle \bar{\psi} | \gamma^{\mu} 
  \frac{1}{\not{\! p} - \not{\! k} - m - \gamma^0 V} \gamma_{\mu} 
  | \psi \rangle
\end{equation}
up to the order of $(Z \alpha)^6$.  
The first step in the evaluation of ${\tilde P}$ is the 
expansion of the matrix 
\begin{displaymath}
M = 
 \gamma^{\mu} \frac{1}{\not{\! p} - \not{\! k} - m - \gamma^0 V} 
\gamma_{\mu}
\end{displaymath}
in powers of the binding field. We denote the denominator 
of the free electron propagator by $D$ ($D = \not{\! p} - \not{\! k} - m$).
Realizing that the binding field $V = - (Z \alpha)^2 \, m / \rho$ carries
two powers of $(Z \alpha)$ (with $\rho = r / a_{\rm Bohr}$), we expand 
the matrix $M$ up to $V^3$, which leads in turn to four matrices,
denoted $M_{i}$,
\begin{eqnarray}
& & M_0 = \gamma^{\mu} \frac{1}{D} \gamma_{\mu}, \quad
M_1 = \gamma^{\mu} \frac{1}{D} \gamma^{0} V \frac{1}{D} \gamma_{\mu}, \quad
M_2 = \gamma^{\mu} \frac{1}{D} \gamma^{0} V \frac{1}{D} \gamma^{0} V
  \frac{1}{D} \gamma_{\mu}, \\
& & M_3 = \gamma^{\mu} \frac{1}{D} \gamma^{0} V \frac{1}{D} \gamma^{0} V
  \frac{1}{D} \gamma^{0} V \frac{1}{D} \gamma_{\mu}. \nonumber
\end{eqnarray}
with $M = M_0 + M_1 + M_2 + M_3 + O((Z \alpha)^7)$.
Defining ${\tilde P}_i = \langle \bar{\psi} | M_i | \psi \rangle$,
we write the element ${\tilde P}$ as the sum
\begin{equation}
{\tilde P} =  {\tilde P}_0 + {\tilde P}_1 + {\tilde P}_2 + {\tilde P}_3 + 
  O((Z \alpha)^7).  
\end{equation}
This expansion corresponds to a division of the initial expression into 
0-,1-,2- and 3-vertex parts.
We then expand each of the matrices $M_i$ into the standard 16 
$\Gamma$ matrices, which form a basis set of $4 \times 4$ matrices.
\begin{equation}
M_i = \sum_{\beta=0}^{15} c_{i,\beta} \, \Gamma^{\beta} \quad \mbox{where} 
\quad
c_{i,\beta} = \frac{1}{4} \, {\rm Tr}(\Gamma_{\beta} M_i).
\end{equation}
The expansion coefficients $c_{i,\beta}$ are rational functions of 
the binding field, the electron and photon energy and  momenta.
They can therefore be  
expanded in powers of $\alpha$, leaving none of the electron 
momentum operators in the denominator. Next, we evaluate 
the matrix elements 
of these operators with the relativistic (Dirac) wave function 
$\psi$. It is a property of $P$ states, which vanish at the 
origin, that up to order $(Z \alpha)^6$,
all of the desired matrix elements are finite.

As an example, we describe here the evaluation of the 
three-vertex matrix 
element ${\tilde P}_3 = \langle \bar{\psi} | M_3 | \psi \rangle$. 
It takes on 
the same  values for both $2P$ states.
Expanding $M_3$ into the 16 $\Gamma$-matrices, we find that 
up to order $(Z \alpha)^6$, all expansion coefficients vanish 
except for the  identity ${\rm Id}$ and  $\gamma^0$-matrices.
The expansion coefficients are explicitly
\begin{equation}
c_{3,{\rm Id}} = 16 V^3 \frac{k^2 - 
k^2 \, \omega - 4 \, \omega + 3 \, \omega^2 + 2}
{ \left( k^2 + 2 \, \omega - \omega^2 \right)^4 }
\equiv b_{3, {\rm Id}} V^3,  
\end{equation}
where $k = | {\bf k} |$ and for simplicity $m=1$, and
\begin{equation}
c_{3,{\gamma^0}} = 2\, V^3 \frac{ k^4 - 8 \, k^2   +
12 \, k^2 \, \omega + 16 \, \omega - 
6 \, k^2 \omega^2 - 12  \, \omega^2 + 4 \, \omega^3 - 
\omega^4 - 8 }
{ \left( k^2 + 2 \, \omega - \omega^2 \right)^4 } \equiv
b_{3, \gamma^0} V^3.
\end{equation}
So up to order $(Z \alpha)^6$, the two $c$-expansion coefficients are
(except for their dependence on $k$ and $\omega$) 
functions of the binding field only.
Thus, the matrix element ${\tilde P}_3$ is given by
\begin{equation}
{\tilde P}_3 = b_{3, {\rm Id}} \langle \bar{\psi} | V^3 | \psi \rangle + 
b_{3, \gamma^0} \langle \bar{\psi} | \gamma^0 V^3 | \psi \rangle.
\end{equation}
The relevant matrix element of the wave function is 
\begin{equation}
\langle \bar{\psi} | V^3 | \psi \rangle =
\langle \bar{\psi} | \gamma^0 V^3 | \psi \rangle = 
- \frac{1}{24} \, (Z \alpha)^6 \, m^3 + O((Z \alpha)^7). 
\end{equation}
where the first equality holds only in the order of $(Z\,\alpha)^6$.
The above matrix elements take on the same values for the
$2P_{1/2}$ and $2P_{3/2}$ states because the radial parts
of both $2P$ states are the same in the non-relativistic limit.

For the other vertex parts, many more terms appear, and the matrix elements
contribute in the lower order also. We give one 
example here, to be evaluated for the 1-vertex part,
\begin{equation}
\langle \bar{\psi} | \gamma^0 {\bf p} \cdot \left(V \, {\bf p} \right) 
| \psi \rangle = - \frac{5}{48} \, (Z \alpha)^4 \, m^3 -
\frac{283}{1152} \, \left( Z \alpha \right)^6 \, m^3 \quad
\mbox{for $2P_{1/2}$}.
\end{equation}
and
\begin{equation}
\langle \bar{\psi} | \gamma^0 {\bf p} \cdot \left(V \, {\bf p} \right) 
| \psi \rangle = - \frac{5}{48} \, (Z \alpha)^4 \, m^3 -
\frac{71}{1152} \, \left( Z \alpha \right)^6 \, m^3 \quad
\mbox{for $2P_{3/2}$}.
\end{equation}
For a more detailed review of the
calculations see \cite{masterthesis}.
Having calculated $\tilde P$,
we subtract the mass-counter-term before integrating with respect
to $k$ and $\omega$. The final $k$ and $\omega$ 
integration is performed in the following way. Those terms which appear
to be ultraviolet divergent are regularized and integrated covariantly 
using
Feynman parameter approach. The remaining terms are
integrated with respect to $k$ by residual integration and with respect
to $\omega$ by changing the integration variable to
\begin{equation}
u = \frac{\sqrt{2 \, m \, \omega - \omega^2} + i \omega}
 {\sqrt{2 \, m \, \omega - \omega^2} - i \omega}.
\end{equation}
This integration procedure is described in details in 
\cite{krpdissertation}. The final results for the high-energy-part 
are (for the definition of $F$ see Eq. (\ref{definitionofF}))
\begin{equation}
F_H(2P_{1/2}) = -\frac{1}{6} + (Z \alpha)^2 
\left[ \frac{4177}{21600} - \frac{103}{180} \ln(2) -
\frac{103}{180} \ln{(\epsilon)} - \frac{2}{9 \epsilon} \right]
\end{equation}
and
\begin{equation}
F_H(2P_{3/2}) = \frac{1}{12} + (Z \alpha)^2 
\left[ \frac{6577}{21600} - \frac{29}{90} \ln(2) -
\frac{29}{90} \ln{(\epsilon)} - \frac{2}{9 \epsilon} \right].
\end{equation}

\section{The low energy part}

In this part we are dealing 
with low energy virtual photons, 
therefore we treat the binding field non-pertubatively. 
Choosing the Coulomb gauge for the photon propagator,
one finds that only the spatial elements of this propagator 
contribute. The $\omega$-integration along $C_L$ is performed 
first, which leads to the following expression for $E_L$,
\begin{equation}
E_L = - e^2 \int_{| {\bf k} | < \epsilon}
\frac{d^3 k}{(2 \pi)^3 \, 2 | {\bf k} |} \, \delta^{T, ij} 
\langle \psi | \alpha^i e^{i \, {\bf k} \, \cdot \, {\bf r}}
\frac{1}{H_D - (E_{\psi} - \omega)}  
\alpha^j e^{-i \, {\bf k} \, \cdot \, {\bf r}} | \psi \rangle
\quad (\omega \equiv | {\bf k} |).
\end{equation}
$H_D$ denotes the 
Dirac-Coulomb-Hamiltonian, $\delta^T$ is the transverse delta function,
and $\alpha^i$ refers to the Dirac $\alpha$-matrices. In the matrix element
\begin{equation}
P^{ij} = 
\langle \psi | \alpha^i e^{i \, {\bf k} \, \cdot \, {\bf r}}
\frac{1}{H_D - (E_{\psi} - \omega)}  
\alpha^j e^{-i \, {\bf k} \, \cdot \, {\bf r}} | \psi \rangle
\end{equation}
we introduce a unitary Foldy-Wouthuysen transformation $U$,
\begin{equation}
P^{ij} = 
\langle U \psi | 
(U \, \alpha^i e^{i \, {\bf k} \, \cdot \, {\bf r}} \, U^{+})
\frac{1}{U \, (H_D - (E_{\psi} - \omega)) \, U^{+}}  
(U \, \alpha^j e^{-i \, {\bf k} \, \cdot \, {\bf r}} \, 
U^{+}) | U \psi \rangle.
\end{equation}
The lower components of the Foldy-Wouthuysen transformed Dirac wave 
function
$\psi$ vanish up to $(Z \alpha)^2$, 
so that we may approximate $| U \psi \rangle$ by
\begin{equation}
| U \psi \rangle = | \phi \rangle + | \delta \phi \rangle \quad
\mbox{with} \quad \langle \phi | \delta \phi \rangle = 0,
\end{equation}
where $| \phi \rangle$ is the nonrelativistic 
(Schr\"{o}dinger-Pauli) wave function,
and $| \delta \phi \rangle$ is the relativistic correction.

We define an operator acting on the spinors as even if it does
not mix upper and lower components of spinors, 
and we call the odd operator odd if it mixes upper and lower
components.  The Foldy-Wouthuysen  Hamiltonian consists of even 
operators only. For the upper left
$2 \times 2$ submatrix of this Hamiltonian, we have the result 
\cite{itzykson} 
\begin{equation}
H_{\rm FW} = U \, (H_D - (E_{\psi} - \omega)) \, U^{+} = m + H_S + \delta 
H,
\end{equation}
where $H_S$ refers to the Schr\"{o}dinger Hamiltonian, and $\delta H$ is
is the relativistic correction,
\begin{equation}
\delta H = - \frac{\left({\bf p}\right)^4}{8 \, m^3}
+ \frac{\pi \alpha}{2 \, m^2} \, \delta({\bf r}) +
\frac{\alpha}{4 \, m^2 \, r^3} \, {\bbox{\sigma}} \, \cdot \, {\bf L}
\end{equation}
Now we turn to the calculation of the Foldy-Wouhuysen transform
of the operators $\alpha^i \exp\left( {\bf k}\, \cdot \, {\bf r}
\right)$. The expression 
$U \, \alpha^i \exp\left(i {\bf k} \, \cdot \, {\bf r}\right) \, U^{+}$
is to be calculated.  
Assuming that $\omega = | {\bf k} |$ is of the order $O((Z \alpha)^2)$,
we may expand the expression
$U \, \alpha^i \, e^{i {\bf k} \, \cdot \, {\bf r}} \, U^{+}$ 
in powers of $(Z \alpha)$.
The result of the calculation is
\begin{eqnarray}
\label{alphairaw}
U \, \alpha^i e^{i {\bf k} \, \cdot \, {\bf r}} \, U^{+} & = &
\alpha^i \left(1 + i \left( {\bf k} \, \cdot {\bf r} \right) -
\frac{1}{2} \left( {\bf k} \, \cdot \, {\bf r} \right)^2 \right) 
- \frac{1}{2 \, m^2} p^i \, \left( \bbox{\alpha} \, \cdot \, \bbox{p} 
\right) \\
& & + \gamma^0 \, 
 \frac{p^i}{m} \left(1 + i \left( {\bf k} \, \cdot {\bf r} \right) -
\frac{1}{2} \left( {\bf k} \, \cdot \, {\bf r} \right)^2 \right)
\nonumber \\
& & - \gamma^0 \frac{1}{2 \, m^3} p^i {\bf p}^2 -
\frac{1}{2 \, m^2} \, \frac{\alpha}{r^3} \, 
  \left( {\bf r} \times \bbox{\Sigma} \right)^i \nonumber \\
& & + \frac{1}{2 \, m} \, \gamma^0 \, 
  \left( {\bf k} \, \cdot \, {\bf r} \right)
     \left( {\bf k} \times \bbox{\Sigma} \right)^i
- \frac{i}{2 \, m} \, \gamma^0 \,
   \left( {\bf k} \times \bbox{\Sigma} \right)^i \nonumber.
\end{eqnarray}
In the limit $\epsilon \to 0$ the odd
operators in the above expression do not contribute to the self energy
in $(Z\,\alpha)^2$ relative order, so one can neglect the odd operators.
It can be shown easily that also the
last term in the above expression (proportional to 
${\bf k} \times \bbox{\Sigma}$) 
does not contribute to the Lamb shift in $(Z \alpha)^2$
relative order for $\epsilon \to 0$. 

Because we can ignore odd operators, and because
the lower components of the Foldy-Wouthuysen transformed wave function
vanish, 
we keep only the upper left $2 \times 2$ submatrix of Eq. 
(\ref{alphairaw}),
and we write 
$U \, \alpha^i \, e^{i {\bf k} \, \cdot \, {\bf r}} U^{+}$ as
\begin{eqnarray}
\label{alphaitransformed}
U \, \alpha^i e^{i {\bf k} \, \cdot \, {\bf r}} \, U^{+} & \simeq &
\frac{p^i}{m} \left(1 + i \left( {\bf k} \, \cdot {\bf r} \right) -
\frac{1}{2} \left( {\bf k} \, \cdot \, {\bf r} \right)^2 \right) \\
& & - \frac{1}{2 \, m^3} p^i {\bf p}^2 -
\frac{1}{2 \, m^2} \, \frac{\alpha}{r^3} \, 
  \left( {\bf r} \times \bbox{\sigma} \right)^i \nonumber \\
& & + \frac{1}{2 \, m} \left( {\bf k} \, \cdot \, {\bf r} \right)
 \left( {\bf k} \times \bbox{\sigma} \right)^i\,, \nonumber
\end{eqnarray}
This can be rewritten as
\begin{equation}
U \, \alpha^i \, e^{i {\bf k} \, \cdot \, {\bf r}} \, U^{+} =
\frac{p^i}{m} \, e^{i {\bf k} \, \cdot \, {\bf r}} + \delta y^i,
\end{equation}
where $\delta y^i$ is of order $(Z \alpha)^3$. It is understood
that the term $\frac{p^i}{m} \, e^{i {\bf k} \, \cdot \, {\bf r}}$ is 
also expanded up to the order $(Z \alpha)^3$. 
Denoting by $E$ the
Schr\"{o}dinger energy ($E = - (Z \alpha)^2 \, m / 8$ for 2P
states) and by $\delta E$ the
first relativistic correction to $E$, we can thus 
write the matrix element $P^{ij}$ as
\begin{equation}
\label{pij}
P^{ij} = 
\langle \phi + \delta \phi | \left[ \frac{p^i}{m} \, 
  e^{i {\bf k} \, \cdot \, {\bf r}} + \delta y^i \right] \,
\frac{1}{H_S - (E - \omega) + \delta H - \delta E} \,
\left[ \frac{p^j}{m} \, 
  e^{- i {\bf k} \, \cdot \, {\bf r}} + \delta y^j \right] 
| \phi + \delta \phi \rangle.
\end{equation}
In this expression, the leading term and the (first)
relativistic corrections can be readily identified. 
Spurious lower order terms are not
present in Eq. (\ref{pij}). By expansion of the denominator
$H_S - (E - \omega) + \delta H - \delta E$ in powers
of $\alpha$, the whole expression can be written in a form 
which involves only the Schr\"{o}dinger-Coulomb-Green's function
\begin{equation}
G(E - \omega) = \frac{1}{H_S - (E - \omega)},
\end{equation}
whose closed-form expression in coordinate space is given in
Eq. (\ref{SCGreensf}). We now define the dimensionless quantity
\begin{equation}
\label{definitionofP}
P = \frac{m}{2} \, \delta^{T, ij} \, P^{ij}.
\end{equation}
Using the symmetry of the $P$-wave functions and Eq. (\ref{pij}),
we easily see that $P$ can be written, up to $(Z \alpha)^2$,
as the sum of the contributions
(\ref{pnd}, \ref{pnq}, \ref{pdeltay}, \ref{pdeltah}, \ref{pdeltae}, 
\ref{pdeltaphi}). 
The leading contribution (the  ``non-relativistic dipole'') is given by
\begin{equation}
\label{pnd}
P_{\rm nd} = \frac{1}{3 m} \, \langle \phi | p^i \, 
\frac{1}{H_S - (E - \omega)} \, p^i | \phi \rangle.
\end{equation}
The evaluation of this matrix element is described here as an example.
For the Schr\"{o}dinger-Coulomb 
propagator, we use the following coordinate-space 
representation \cite{swainsondrake},
\begin{equation}
\label{SCGreensf}
G({\bf r}_1, {\bf r}_2, E - \omega) = \sum_{l,m} \,
g_l({\bf r}_1, {\bf r}_2, \nu) \,
Y_{l,m} \left(\hat{\bf r}_1\right) \, 
Y_{l,m}^{*} \left(\hat{\bf r}_2\right),
\end{equation}
with $E - \omega \equiv - \alpha^2 \, m / (2 \nu^2)$.
\begin{equation}
\label{gl}
g_l(r_1, r_2, \nu) = \frac{4 m}{a \nu} 
\left( \frac{2 r_1}{a \nu} \right)^l \, 
\left( \frac{2 r_1}{a \nu} \right)^l \,
e^{- (r_1 + r_2)/(a \nu) } 
\sum_{k=0}^{\infty} 
\frac{L_k^{2 l + 1}\left(\frac{2 r_1}{a \nu} \right) \, 
L_k^{2 l + 1}\left(\frac{2 r_2}{a \nu} \right)}
{(k+1)_{2l+1} \, (l + 1 + k - \nu)},
\nonumber
\end{equation}
where $a = a_{\rm Bohr} = 1 / (\alpha m)$, 
and $(k)_c$ is the Pochhammer symbol. 
The evaluation of eq. (\ref{pnd}) 
proceeds in the following steps: The angular integration
is performed first. Secondly, the remaining integrals over $r_1$ and $r_2$ 
are evaluated using the formula (see e.g. \cite{buchholz}), 
\begin{eqnarray}
\int_0^\infty dt \,e^{-s t} \,t^{\gamma-1} \,L_n^\mu (t)  = 
\frac{\Gamma(\gamma) \Gamma(n + \mu + 1)}
{n! \,\Gamma(\mu + 1)}
s^{-\gamma} \,{}_2F_1\Bigl(-n, \gamma, 1+\mu;\frac{1}{s}\Bigr)\,.
\end{eqnarray}
The following formula is useful for carrying out the summation 
with respect to $k$ \cite{bateman},
\begin{equation}
\label{sumformula}
\sum_{n=0}^{\infty}
\frac{\Gamma(n+\lambda)}{n!} \, s^n \,{_2}F_1(-n,b;c;z) = 
\Gamma(\lambda) \, (1-s)^{-\lambda} \, 
{_2}F_1\Bigl(\lambda,b;c;-\frac{s\, z}{1-s}\Bigr).
\end{equation}
The summations lead to hypergeometric functions in the result,
\begin{eqnarray}
P_{\rm nd}(t) &=& 
\frac{ 2 t^2 \left( 3 - 6\,t - 3\,{t^2} + 12\,{t^3} + 29\,{t^4} + 
       122\,{t^5} -  413\,{t^6} \right) } 
    {9\, \left( 1 - t \right)^5 \, \left( 1 + t \right) ^3} +\nonumber \\
& & \frac{256\,t^7 \, \left( -3 + 11\,t^2 \right) } 
    {9\,\left( 1 - t \right)^5 \, \left( 1 + t \right)^5} \, 
    {_2}F_1 \left(1, - 2t; 1- 2t; \left( \frac{1-t}{1+t} \right)^2 
            \right)
\end{eqnarray}
where
\begin{equation}
t \equiv \frac{\sqrt{-2 \, m \, E}}{\sqrt{-2 \, m \, (E - \omega)}} = 
  \frac{1}{2} \nu.
\end{equation}
In this expression the terms that gives divergent in $\epsilon$ terms
are separated out of the hypergeometric function, so the could be easily
integrated out. The other contributions to $P$ (for definition fo $P$ see
eq. (\ref{definitionofP})) are
\begin{itemize}
\item the non-relativistic quadrupole,
\begin{equation}
\label{pnq}
P_{\rm nq} = \frac{1}{3 m} \, \langle \phi | p^i \, 
e^{ i {\bf k} \, \cdot \, {\bf r}} \, 
\frac{1}{H_S - (E - \omega)} \, p^i \, 
e^{ - i {\bf k} \, \cdot \, {\bf r}} | \phi \rangle - P_{\rm nd},
\end{equation}
\item the corrections to the current $\alpha^i$ from the
Foldy-Wouthuysen transformation,
\begin{equation}
\label{pdeltay}
P_{\delta y} = \delta^{T, ij} \, 
\langle \phi | \delta y^i \, 
\frac{1}{H_S - (E - \omega)} \, p^j \, 
e^{ - i {\bf k} \, \cdot \, {\bf r}} | \phi \rangle,
\end{equation}
\item the contribution due to the relativistic Hamiltonian,
\begin{equation}
\label{pdeltah}
P_{\delta H} = - \frac{1}{3 m} \, \langle \phi | p^i \, 
\frac{1}{H_S - (E - \omega)} \, \delta H \, 
\frac{1}{H_S - (E - \omega)} \, p^i | \phi \rangle,
\end{equation}
\item the contribution due to the relativistic correction to the
energy,
\begin{equation}
\label{pdeltae}
P_{\delta E} = \frac{1}{3 m} \, \langle \phi | p^i \, 
\frac{1}{H_S - (E - \omega)} \, \delta E \, 
\frac{1}{H_S - (E - \omega)} \, p^i | \phi \rangle,
\end{equation}
\item and due to the relativistic correction
to the wave function,
\begin{equation}
\label{pdeltaphi}
P_{\delta \phi} = \frac{2}{3 m} \, \langle \delta \phi | p^i \, 
\frac{1}{H_S - (E - \omega)} \, p^i | \phi \rangle.
\end{equation}
\end{itemize}
For almost all of the matrix elements we use the coordinate-space 
representation of the Schr\"{o}dinger-Coulomb propagator given in Eq.
(\ref{SCGreensf}). 
There are two exceptions: For the non-relativististic quadrupole,
we use Schwinger's momentum space representation and carry out the
calculation in momentum space. A rather involved contribution is
\begin{equation}
P_{\delta H} = - \frac{1}{3 m} \, \langle \phi | p^i \, G(E - \omega) \,  
\left[ - \frac{\left({\bf p}\right)^4}{8 \, m^3}
+ \frac{\pi \alpha}{2 \, m^2} \, \delta({\bf r}) +
\frac{\alpha}{4 \, m^2 \, r^3} \, {\bbox{\sigma}} \, \cdot \, {\bf L}
\right] \, G(E - \omega) \, p^i | \phi \rangle.
\end{equation}
where $G(E - \omega) = 1 / (H_S - (E - \omega))$. The form 
of $\delta H$ implies a natural separation of $P_{\delta H}$
into three terms,
\begin{equation}
P_{\delta H} = P_{p^4} + P_{\delta} + P_{L \cdot S}.
\end{equation}
For $P_{\delta}$, 
\begin{equation}
P_{\delta} = - \frac{1}{3 m} \, \langle \phi | p^i \, G(E - \omega) \, 
\left[\frac{\pi \alpha}{2 \, m^2} \, \delta({\bf r}) \right] 
\, G(E - \omega) \, p^i | \phi \rangle,
\end{equation}
which involves the zitterbewegungs-term 
(proportional to the $\delta$-function), we
use a coordinate-space representation of the
Schr\"{o}dinger-Coulomb propagator involving Whittaker functions
(this representation is also to be found in \cite{swainsondrake}). 
The result for $P_{\delta}(t)$ is
\begin{equation}
P_{\delta}(t) = 
-\frac{\alpha^2}{27} \,
  \frac{t^4\, \left( -3 + 4\,t + 7\,{t^2} - 8\,t\,F_2(t) \right)^2}
 {\left( t^2 - 1 \right)^4}
\end{equation}
where
\begin{equation}
F_2(t) = {_2}F_1\left(1,-2\,t,1-2\,t,\frac{t-1}{t+1}\right).
\end{equation}
Both terms $P_{p^4}$ and $P_{L \cdot S}$,
\begin{eqnarray}
P_{p^4} & = &  - \frac{1}{3 m} \, \langle \phi | p^i \, G(E - \omega) \, 
\left[ - \frac{\left({\bf p}\right)^4}{8 \, m^3}  \right] 
\, G(E - \omega) \, p^i | \phi \rangle\\
P_{L \cdot S} & = &  - \frac{1}{3 m} \, \langle \phi | p^i \, G(E - 
\omega) \, 
\left[ \frac{\alpha}{4 \, m^2 \, r^3} \, {\bbox{\sigma}} \, 
\cdot \, {\bf L} \right] 
\, G(E - \omega) \, p^i | \phi \rangle,
\end{eqnarray}
involve two propagators $G(E-\omega)$.
We use the Schr\"{o}dinger equation and the identity
\begin{equation}
[H_S - (E - \omega), \, \frac{1}{r} \, \frac{\partial}{\partial r} \, r] =
\frac{{\bf L}^2}{m \, r^3} - \frac{Z \alpha}{r^2}.
\end{equation}
to rewrite them to the form that contain only
one  propagator with modified parameters. Namely,
to the desired order in $(Z \alpha)$,
the expression with two propagators can be replaced by an expression with
just one propagator, in which an $(Z \alpha)^2$-correction is
added to the angular momentum parameter $l$ or to the 
fine structure constant $\alpha$
in the radial part of the Schr\"{o}dinger-Coulomb propagator
as given in Eq. (\ref{SCGreensf}).
For the $P_{p^4}$ and $P_{L \cdot S}$ contributions,
many more terms appear in the calculation, and 
derivatives of the hypergeometric functions with respect to parameters 
have to be evaluated. 
The result consists of terms involving elementary functions and 
hypergeometric functions only, and other terms which involve slightly
more complex functions. Some of the summations give rise
to the Lerch transcendent $\Phi$. Summations of the form
\begin{equation}
\sum_{k=0}^{\infty} k^n {\xi^k}\,
  \frac{\partial}{\partial b}\, {_2}F_1(-k,b,c,z).
\end{equation}
can be evaluated with the help of 
Eq. (\ref{sumformula}), for  more details see \cite{masterthesis}.
Although we do not describe the calculations in detail, we stress
that the summation with respect to the $k$-index is 
the decisive point in the calculation. 
In general, a sensible use of contiguous relations is necessary to simplify
the result of any of the summations. Symbolic procedures
were written to accomplish this. Through the compartmentalization of the
calculation achieved by the Foldy-Wouthuysen transformation,
it has been possible to keep the average length of intermediate expressions
below 1000 terms.

The contribution to $E_L$ due to the $\delta E_{\rm SE}$  is given by
\begin{equation}
E_L = -\frac{2 \, \alpha}{\pi \, m} 
\int_0^{\epsilon} d \omega \, \omega \, P(\omega).
\end{equation}
Changing the integration variable to $t$, we have
\begin{equation}
F = -\frac{1}{2} \, \int_{t_{\epsilon}}^{1} dt \, 
\frac{1-t^2}{t^5} \, P(t).
\end{equation}
The $P$-terms are integrated with respect to $t$ by the following 
procedure. Terms which give a divergence
for $\epsilon \to 0$ are extracted from the integrand. 
The extraction can be achieved by a suitable expansion 
in the argument of the hypergeometric function(s) 
which appear in $P(t)$.
The extracted terms consist of elementary 
functions of $t$ only, so they can be integrated 
analytically. After integration, the terms are first
expanded in $(Z \alpha)$ up to $(Z \alpha)^2$, then in $\epsilon$
up to $\epsilon^0$. 
The remaining part, which involves hypergeometric functions,
is integrated numerically with respect to $t$ by the Gaussian method. 

The $t$-integration leads to $F$-terms which we name according
to the $P$-terms $F_{\rm nd}$, 
$F_{\rm nq}$, $F_{\delta y}$, $F_{\delta H}$,
$F_{\delta E}$ and $F_{\delta \phi}$. The $F_{\rm nd}$-term, 
which is the same for both $2P$-states, is given by
\begin{equation}
F_{\rm nd} = - \frac{4}{3} \ln k_0 (2P) + \frac{2}{9} \,
\frac{(Z \alpha)^2}{\epsilon}.
\end{equation}
We have recovered the first 9 digits of the Bethe logarithm with
our (Gaussian) integration procedure (the value for
the Bethe logarithm given in \cite{sapirstein} is
$\ln k_0(2P) = -0.0300167089(3)$).
The $F_{\rm nd}$-term has, for $\epsilon \to 0$, a 
divergence of $+ 2/9 (Z \alpha)^2 / \epsilon$, which cancels the 
corresponding divergence in the high energy part. All other $F$-terms 
produce logarithmic divergences in $(Z \alpha)^2 \ln(\epsilon)$
(see Table \ref{table1}).
The results for the low-energy parts of the $2P$-states are
\begin{equation}
F_L(2P_{1/2}) = -\frac{4}{3} \, \ln k_0 (2P) + 
\left(Z \alpha \right)^2 \left[ -0.79565(1) + 
\frac{103}{180} \ln\left( (Z \alpha)^{-2} \right) +
\frac{103}{180} \ln\left( \epsilon \right) +
\frac{2}{9 \, \epsilon} \right]
\end{equation}
and
\begin{equation}
F_L(2P_{3/2}) = -\frac{4}{3} \, \ln k_0 (2P) + 
\left(Z \alpha \right)^2 \left[ -0.58452(1) + 
\frac{29}{90} \ln\left( (Z \alpha)^{-2} \right) +
\frac{29}{90} \ln\left( \epsilon \right) +
\frac{2}{9 \, \epsilon} \right].
\end{equation}
The divergence in 
$1/\epsilon$ and in $\ln(\epsilon)$ cancels out when the low- and 
high-energy-parts are added. The results for the $F$-factors 
(sum of low-energy-part and high-energy-part) are:
\begin{equation}
F(2P_{1/2}) = -\frac{1}{12} - \frac{4}{3} \, \ln k_0 (2P) + 
\left(Z \alpha \right)^2 \left[ -0.99891(1) + 
\frac{103}{180} \ln\left( (Z \alpha)^{-2} \right) \right]
\end{equation}
for the $2P_{1/2}$-state and
\begin{equation}
F(2P_{3/2}) = \frac{1}{6} - \frac{4}{3} \, \ln k_0 (2P) + 
\left(Z \alpha \right)^2 \left[ -0.50337(1) + 
\frac{29}{90} \ln\left( (Z \alpha)^{-2} \right) \right]
\end{equation}
for the $2P_{3/2}$-state. The $A_{60}$ coefficients are given by
\begin{equation}
A_{60}(2P_{1/2}) = -0.99891(1)
\end{equation}
and
\begin{equation}
A_{60}(2P_{3/2}) = -0.50337(1).
\end{equation}
The last digit is the cumulated inaccuracy of the numerical
integrations. The values for the $A_{40}$ and $A_{61}$ coefficients 
are in agreement with known results \cite{sapirstein}.

These results can be compared to those obtained by P. Mohr \cite{mohr3}
by extrapolation of his numerical data for higher Z,
\begin{equation}
G_{SE}(2) = -0.96(4),\;\;\;G_{SE}(1) = -0.98(4) \quad \mbox{for 
$2P_{1/2}$},
\end{equation}
and
\begin{equation}
G_{SE}(2) = -0.46(2),\;\;\;G_{SE}(1) = -0.48(2) \quad \mbox{for 
$2P_{3/2}$},
\end{equation}
where the function $G_{\rm SE}(Z)$ for $2P$-states is defined by
\begin{equation}
F = A_{40} + (Z \alpha)^2 \left[ A_{61} \, \ln\left( (Z \alpha)^{-2} 
\right)
 + G_{\rm SE}(Z) \right].
\end{equation}
Because $G_{\rm SE}(Z=0) = A_{60}$, these values are clearly in very good
agreement with the results of our analytical calculation.
Using P. Mohr's numerical data \cite{mohr}, we have obtained the 
following estimates for higher order terms summarized by $G_{\rm SE,7}$  
\begin{equation}
F = A_{40} + (Z \alpha)^2 \left[ A_{60} + A_{61} \,
 \ln\left( (Z \alpha)^{-2} \right)
  + (Z \alpha) \, G_{\rm SE,7}(Z) \right],
\end{equation}
\begin{equation}
\label{GSEseven}
G_{\rm SE,7}(2P_{1/2},Z=1) = 3.1(5) \quad \mbox{and} \quad
G_{\rm SE,7}(2P_{3/2},Z=1) = 2.3(5). 
\end{equation}

One of the most important aspects of rather lengthy calculations 
such as those presented here is to avoid errors. The
result has been checked in many ways. Except for checking the values
of the terms divergent in $\epsilon$, it was also checked
the value of each $P$-contribution as $\omega \to 0$. It can be shown 
easily that the sum of all contributions to the matrix element
$P$ in the low-energy part must vanish
in the limit $\omega \to 0$. 
Care must be taken when checking the sum, because after 
the Foldy-Wouthuysen transformation, hidden 
terms are introduced which do not contribute to the Lamb shift, but 
contribute in the limit $\omega \to 0$. The hidden terms originate from
the odd operators in Eq. (\ref{alphairaw}). Taking into account these 
terms, the sum vanishes for both states. 

\section{Other contributions to the Lamb shift}

For the Lamb shift ${\cal L}$, 
we use the implicit definition
\begin{equation}
\label{defElamb}
E = m_r \left[ f(n,j)-1 \right] - \frac{m_r^2}{2 (m + m_N)}
\left[ f(n,j) - 1 \right]^2 + {\cal L} +
E_{\rm hfs},
\end{equation}
where $E$ is the energy level of the two-body-system 
and $f(n,j)$ is the dimensionless Dirac energy, $m$
is the electron mass, $m_r$ is the reduced mass of the system 
and $m_N$ is the nuclear mass.

For the final evaluation of the Lamb shift
the following contributions are added:
\begin{enumerate}
\item One-loop self energy. The coefficients are
presented in this work. For the determination
of the Lamb shift the reduced mass dependence of 
the terms has to be restored. The relevant formulae are
given in \cite{sapirstein}. For example, the $A_{60}$ 
have a reduced mass dependence 
of $(m_r/m)^3$. We use Eq. (\ref{GSEseven}) to estimate the theoretical
uncertainty from the one--loop contribution.
\item Vacuum polarization correction. It
enters for $P$-states in higher order (for the formulae see 
\cite{sapirstein}, p. 570).
\item Two-loop contributions due to the anomalous magnetic moment 
\cite{kinoshita2}. It is given in analogy to the one-loop contribution as 
\begin{equation}
\delta E_{\rm 2-loop} = \left(\frac{\alpha}{\pi}\right)^2 \, m \,
\frac{(Z \alpha)^4}{n^3} \left[ B_{40} + \dots \right]
\end{equation}  
where the $B$-coefficients are labeled in analogy  to the 
$A$-coefficients for the one-loop self energy.
The $B_{40}$ coefficient is due to the anomalous magnetic moment
of the electron. It is given as 
\begin{equation}
B_{40} = \frac{C_{jl}}{2(2l + 1)} \,
\left[ \frac{197}{72} +\frac{\pi^2}{6} - \pi^2 \ln 2 + 
\frac{3}{2} \zeta(3) \right] \, \left(\frac{m_r}{m} \right)^2,
\end{equation}
where $C_{jl} = 2(j-l)/(j+1/2)$. 
\item Two loop contributions in higher order. Recently, the logarithmic 
term 
\begin{equation}
B_{62} =  \left[ \frac{4}{27} \, 
\frac{n^2 - 1}{n^2} \, \ln^2\left((Z \alpha)^{-2}\right) \right] \,
\left(\frac{m_r}{m} \right)^3,
\end{equation}
has been calculated in
\cite{karshenboim}. The $B_{62}$ term, which is
enlarged by the logarithm, probably dominates the contributions to the 
two-loop self energy in higher order. So the result may also be used 
to estimate the theoretical uncertainty of the two--loop contribution,
coming mainly from the unknown $B_{61}$ coefficient.
It is taken to be half the contribution from $B_{62}$.
\item Three-loop self energy as given by the anomalous 
magnetic moment \cite{kinoshita2}.
\begin{equation}
\delta E_{\rm 3-loop} = \left(\frac{\alpha}{\pi}\right)^3 \, m \,
\frac{(Z \alpha)^4}{n^3} \left[ C_{40} + \dots \right]
\end{equation}
where
\begin{equation}
C_{40} = \left[ 2 \, \frac{C_{jl}}{2(2 \, l +1)} \, 1.17611(1) \right] \,
\left(\frac{m_r}{m} \right)^2.
\end{equation}
\item The additional reduced mass dependence 
of order $(m_r/m_N)^2 \, (Z \alpha)^4$ \cite{sapirstein},
which we will refer to as the $(Z \alpha)^4$ recoil contribution,
\begin{equation}
\delta E_{\rm rec,4} = \frac{(Z \alpha)^4}{2 \, n^3} \,
\frac{m_r^3}{m_N^2}\,
\left( \frac{1}{j + 1/2} - \frac{1}{l + 1/2} \right) \, 
\left(1 - \delta_{l0} \right),
\end{equation}
\item The Salpeter correction (relativistic recoil) in order
$(Z \alpha)^5$ as given in \cite{sapirstein}. The formula is for $P$-states
\begin{equation}
\delta E_{\rm rec,5} = \frac{m_r^3}{m\,m_N}\,
\frac{(Z \alpha)^5}{\pi \, n^3} \,
\left( - \frac{8}{3} \ln k_0 (n) - 
\frac{7}{3} \frac{1}{l(l+1)(2\,l+1)} \right).
\end{equation}
\item Relativistic recoil corrections in the order of
$(Z \alpha)^6 \, m_r/m_N$,
\begin{equation}
\delta E_{\rm rec,6} = \frac{m^2}{m_N} \, 
  (Z \alpha)^6 \,
\left[ \frac{1}{2} \langle{\phi} | \frac{{\bf L}^2}{r^4} | \phi \rangle 
\right].
\end{equation}
The formula for P-states has been first calculated in
\cite{golosov}. This general form has been obtained by us.
\end{enumerate}
The above contributions are listed in table \ref{lamb2p} for the
$2P$ states.
\section{Results and Conclusions}

The new theoretical values for the Lamb shifts of the 
$2P_{1/2}$ and $2P_{3/2}$ states are
\begin{equation}
{\cal L}(2P_{1/2}) = -12835.99(8) \, {\rm kHz}
\end{equation}
and
\begin{equation}
{\cal L}(2P_{3/2}) = 12517.46(8) \, {\rm kHz}.
\end{equation}
From the values of the $2P$ Lamb shifts, the fine structure
can be determined. It turns out that the limiting factor in 
the uncertatinty is the experimental value of the fine structure
constant $\alpha$. Using a value of \cite{cohen} (1987)

\begin{equation}
\alpha^{-1} = 137.0359895(61) \quad \mbox{(44 ppb)},
\end{equation}
the fine structure can be determined as
\begin{equation}
E(2P_{3/2}) - E(2P_{1/2}) = 10969043(1) \, {\rm kHz}.
\end{equation}
With the most recent and most precise value of $\alpha$ available 
\cite{kinoshita} (1995),
\begin{equation}
\alpha^{-1} = 137.03599944(57) \quad \mbox{(4.2 ppb)},
\end{equation}
we obtain a value of 
\begin{equation}
E(2P_{3/2}) - E(2P_{1/2}) = 10969041.52(9)(8) \, {\rm kHz},
\end{equation}
where the first error originates from the uncertainty in $\alpha$
and the second from the uncertainty in the Lamb shift difference.
Our result for the fine structure
disagrees with that used by Hagley and Pipkin in \cite{hagley}
for the determination of $L(2S-2P_{1/2})$.
Therefore their result of $L(2S-2P_{1/2}) = 1057839(12)$
is to be modified and according to our calculation it should be
\begin{equation}
L(2S-2P_{1/2}) = 1057842(12) \, {\rm kHz}.
\end{equation} 

Precise theoretical predictions for $P$-states could be used to 
compare two different kind of measurements of Lamb shifts in the hydrogen. 
One is the classic $2S_{1/2}$-$2P_{1/2}$ Lamb shift
measured by several groups \cite{lundeen}, \cite{palchikov},
\cite{hagley}, and the second is the combined Lamb shift
${\cal L}(4S - 2S) - \frac{1}{4} \, {\cal L}(2S-1S)$ 
as measured by the H\"{a}nsch group (for a review see \cite{edirne}).
The experimental value of 2S Lamb shift can be extracted from 
E($2S$-$2P_{1/2}$)
having the precise value for $2P_{1/2}$ Lamb shift, and can also be
determined from the combined Lamb shift through the formula
\begin{equation}
{\cal L}(2S) = \frac{8}{7}
\left[ \left({\cal L}(4S) - \frac{5}{4} {\cal L}(2S) + 
{\cal L}(1S)\right)_{\rm exp} - 
\left({\cal L}(4S) - \frac{17}{8} {\cal L}(2S) + 
{\cal L}(1S)\right)_{\rm theo} \right], 
\end{equation} 
where the subscript {\em exp} denotes experimental, and the subscript 
{\em theo} denotes theoretical values. This {\em theo} combination has the 
property 
that terms scaling $1/n^3$ cancel out, which means that almost
all QED effects do not contribute, and therefore
the quantity can be precisely determined.
Such a comparison of completely different experimental
techniques is an interesting and valuable test of 
high precision experiments.

The method of calculation presented in this paper could be directly applied
for the evaluation of Lamb shifts and the fine structure
in two electron systems, for example in helium or positronium.
It was a purpose of this method to use only a 
Schr\"{o}dinger-Coulomb propagator, and relativistic effects 
are incorporated through the Foldy-Wouthuysen transformation.
This method clearly separates out the lower and the higher order
terms, and expresses the energy shift
through the matrix elements of nonrelativistic operators.

\section*{Acknowledgments}

This work was done while one of us (K. P.) was a guest 
scientist at the Max-Planck-Institute for Quantum Optics.
The authors would like to thank T. W. H\"{a}nsch for 
hospitaliy, encouragement and stimulation.
We are very grateful to P. Mohr for supplying extrapolation data
of his 1992 calculations, and to M. Weitz and A. Weis for 
carefully reading the manuscript.
(U. J.) would also like to thank H. Kalf for helpful discussions 
with respect to the treatment of hypergeometric functions.\\[3ex]
{\em Note added (2000): The analytic results for higher-order
binding corrections to the Lamb shift of $2P_{1/2}$ 
and $2P_{3/2}$--states (in particular, the $A_{60}$--coefficient)
have recently been confirmed
by an improved  numerical calculation in the range of low nuclear charge
numbers $Z = 1$--$5$. For details see the e-print {\tt physics/0009090}.}

%
%

\begin{table}[htb]
\begin{center}
\begin{tabular}{c|c|c} 
\rule[-3mm]{0mm}{8mm}
contribution & $2P_{1/2}$ &  $2P_{3/2}$ \\ \hline
\rule[-3mm]{0mm}{8mm}
$F_{\rm nq}$ & 
$-1.201150(1) + 49/90 \ln\left(\epsilon/(Z \alpha)^2\right) $
& $-1.201150(1) + 49/90\ln\left(\epsilon/(Z \alpha)^2\right)$ \\  
\rule[-3mm]{0mm}{8mm}
$F_{\delta y}$ & 
$0.791493(1) - 2/9\ln\left(\epsilon/(Z \alpha)^2\right)$ 
 & $0.531475(1) - 2/9\ln\left(\epsilon/(Z \alpha)^2\right)$ \\ 
\rule[-3mm]{0mm}{8mm}
$F_{\delta H}$ & 
$0.322389(1) - 47/288\ln\left(\epsilon/(Z \alpha)^2\right)$ 
& $0.293749(1) - 35/288\ln\left(\epsilon/(Z \alpha)^2\right)$ \\ 
\rule[-3mm]{0mm}{8mm}
$F_{\delta E}$ &
$0.040095(1) + 5/96 \ln\left(\epsilon/(Z \alpha)^2\right)$ 
 & $ 0.008019(1) + 1/96\ln\left(\epsilon/(Z \alpha)^2\right)$ \\  
\rule[-3mm]{0mm}{8mm}
$F_{\delta \phi}$ & 
$-0.748478(1) + 13/36\ln\left(\epsilon/(Z \alpha)^2\right)$
 & $ -0.216612(1) + 1/96\ln\left(\epsilon/(Z \alpha)^2\right)$ \\ \hline
\rule[-4mm]{0mm}{10mm}
sum &
$-0.79565(1) + 103/180\ln\left(\epsilon/(Z \alpha)^2\right)$ 
 & $-0.58452(1) + 29/90\ln\left(\epsilon/(Z \alpha)^2\right)$ 
\end{tabular}
\end{center}
\caption{\label{table1} Contributions of relative order 
$(Z \alpha)^2$ to the low energy part $F_L$ for the 
$2P_{1/2}$ and $2P_{3/2}$ states }
\end{table}

%
%

\begin{table}
\begin{center}
\begin{tabular}{c|r|r} 
\rule[-3mm]{0mm}{8mm}
contribution & $2P_{1/2}$ in kHz & $2P_{3/2}$ in kHz \\ \hline
\rule[-3mm]{0mm}{8mm}
one-loop self-energy & $-12846.92(2)$ & $12547.95(2)$ \\ 
\rule[-3mm]{0mm}{8mm}
two-loop self-energy & $25.98(7)$ & $-12.79(7)$ \\ 
\rule[-3mm]{0mm}{8mm}
three-loop self-energy & $-0.21$ & $0.10$ \\ 
\rule[-3mm]{0mm}{8mm}
vacuum polarization & $-0.35$ & $-0.08$ \\ 
\rule[-3mm]{0mm}{8mm}
$(Z \alpha)^4$ recoil & $2.16$ & $-1.08$ \\ 
\rule[-3mm]{0mm}{8mm}
$(Z \alpha)^5$ recoil & $-17.08$ & $-17.08$ \\ 
\rule[-3mm]{0mm}{8mm}
$(Z \alpha)^6$ recoil & $0.42$ & $0.42$ \\ \hline
\rule[-3mm]{0mm}{8mm}
sum for $2P_{1/2}$ & $-12835.99(8)$ & $12517.46(8)$  
\end{tabular}
\end{center}
\caption{\label{lamb2p} Contributions to the Lamb shift in kHz for the 
$2P_{1/2}$ and $2P_{3/2}$ states. Estimates of the contributions of 
uncalculated higher order terms are given in the text.
Where no uncertainties are specified, they are negligible at the current
level of precision.}
\end{table}

%
%
%

\end{document}